\documentclass[a4paper]{article}

\usepackage{ra}

\title{\Large Technocracy, pseudoscience \& performative compliance: \\
the risks of privacy risk assessments\\ \vspace{0.3cm}
 \large Lessons from NIST's Privacy Risk Assessment Methodology\footnote{
\textbf{Working draft.  A version of this paper was presented at the 16th International Conference on Computers, Privacy \& Data Protection, 
May 24--26, 2023 in Brussels (Belgium). }
 }}

\author{Ero Balsa}

\date{
	\today
}

\begin{document}
	\maketitle
	
\begin{abstract}
Privacy risk assessments have been touted as an objective, principled way to encourage organizations to implement privacy-by-design.
They are central to a new regulatory model of collaborative governance, as embodied by the GDPR.
However, existing guidelines and methods remain vague, and there is little empirical evidence on privacy harms. In this paper we conduct a close analysis of US NIST’s Privacy Risk Assessment Methodology, 
highlighting multiple sites of discretion that create countless opportunities 
for adversarial organizations to engage in performative compliance. 
Our~analysis shows that the premises on which the success of privacy risk assessments depends do not hold,
particularly in regard to organizations’ incentives and regulators auditing capabilities. 
We~highlight the limitations and pitfalls of what is essentially a utilitarian and technocratic approach, 
leading us to discuss alternatives and a realignment of our policy and research objectives. 
\end{abstract}
	
% intro.tex

\section{Introduction}

% The rise of privacy risk assessments as a policy instrument :: The promises of collaborative governance
\Glspl{pra} have steadily gained prominence as an instrument to help and encourage 
organizations to develop privacy-preserving systems and services.
Touted as a much needed departure from the rigid command-and-control regulatory model,
\glspl{pra} are said to embody a more agile and dynamic approach to privacy regulation.
This model, variously referred to as meta-regulation or collaborative governance,
seeks to leverage knowledge and expertise from organizations, as well as push responsibility onto them, 
encouraging cooperation and trust 
with the regulator in a light supervisory role~\cite{binns2017data,kaminski2018binary,kaminski2020algorithmic} 
At the same time,  by encouraging organizations to assess privacy risks before 
they implement and deploy their services,
\glspl{pra} are considered a key tool in implementing privacy-by-design~\cite{cavoukian2009privacy,spiekermann2012challenges}. 
%
% GDPR and NIST
Indeed, \glspl{pra} have recently gained renewed support across both sides of the Atlantic, 
with the EU enshrining risk assessments in the GDPR, 
and the US's NIST proposing risk assessment as the key tool in their privacy engineering program~\cite{gdpr,gellert2018understanding,gellert2020risk,nistPF,nistir8062}.

However, in spite of the institutional endorsement, the actual benefits of adopting \glspl{pra} remains unclear. 
Recent empirical work as well as the constant trickle of privacy breaches, scandals and litigation, 
suggests that \glspl{pia} may not be as effective as expected~\cite{schechner2022meta,waldman2021industry}. 
%
% Long tradition of PIAs :: the conditions for success 
% Premises around co-/collaborative governance :: incentives (orgs due diligence + accountability from regulator)
Legal scholars and political scientists have long theorized about the necessary conditions for \glspl{pia} to succeed~\cite{bennett2020revisiting,wright2012privacy} 
Factors such as \emph{transparency}, 
\emph{accountability}, 
and a systematic, principled approach to risk assessment, 
e.g.~recital 76 of the GDPR specifies that \emph{``[r]isk should be evaluated on the basis of an objective assessment''}~\cite{gdpr}.

% The utopian horizon  + Increasing embodiment of a technocentric approach + embedding into privacy engineering
This emphasis in objectivity further hints at a technocratic approach to privacy regulation that pervades \gls{pia} scholarship~\cite{gellert2020risk}. 
Early on,  \gls{pia} proponents lamented the lack of standardized guidelines and trained professionals, 
as well as the vagueness and arbitrariness of existing \gls{pia}~\cite{raab1996taking,wright2012privacy}. 
Privacy scholars argued that a systematic and (somewhat) objective approach to risk assessment 
could however turn the tide and not only help organizations manage privacy risks, 
but render \glspl{pia} more successful overall~\cite{raab1996taking}.
Yet this ambition has also failed to materialize.  Privacy harms remain elusive to model and quantify.\footnote{
While recent scholarship and discourse has increasingly recognized and widened the scope of privacy harms,
privacy risk remains absent from the mainstream risk analysis literature~\cite{calo2011boundaries,citron2022privacy,solove2017risk}. 
}
We do not seem today any closer to having something even resembling a \emph{science} of privacy risks 
or \emph{actuarial models} for privacy than 30 years ago.
\gls{pia} has remained more of an art than a science,  learned by craft and practice rather than formal methods. 

This paper revisits and re-examines \gls{pra} as a tool in the regulatory arsenal, 
providing the following contributions. 
Firstly, we question whether \glspl{pra} 
can, today, deliver meaningful results in terms of increased privacy protections.
We investigate the extent to which \glspl{pra} 
(1) may be weaponized by \emph{adversarial} organizations as a tool of \emph{performative compliance}~\cite{waldman2021industry},
and (2) are amenable to the kind of systematic audit processes that collaborative governance requires. 
We observe that existing guidelines 
introduce (to borrow Green's terminology) countless \emph{sites of discretion} that organizations can weaponize 
to escape meaningful privacy intervention and rendering them, as a result,  hard to audit~\cite{green2020false}.

Secondly,  we examine the role of risk assessments within the field of privacy engineering~\cite{gurses2016privacy}.
Privacy engineers have become adept at minimizing privacy risk through privacy enhancing technologies~\cite{gurses2015engineering}.
However,  privacy engineering does not support---cannot support---the broader scope of risk assessment 
that methodologies such as NIST's \Gls{pram} and regulatory efforts such as the GDPR promote.
When it comes to assessing the legitimacy of a service or functionality, privacy engineering is mute.

Thirdly,  we resort to Nissenbaum's theory of \Gls{ci} to show that \glspl{pra} 
have a disruptive potential insofar as they can weaponize the lack of empirical
evidence on privacy harms to upend existing norms and expectations around privacy~\cite{nissenbaum2009privacy}. 
We further argue how \gls{ci} is in fact a \gls{pra} of sorts,
even if devoid of the pseudoscientific claims and technocratic ideals that \gls{pra} embody, 
and may offer hints to a better deliberation framework to guide privacy design. 

Furthermore, to support and illustrate the points above, we provide a step-by-step analysis of NIST's \Gls{pram},
While NIST's \gls{pram} is just one among many guidelines and methods of \gls{pra}, 
NIST's normative authority as a standard-setting agency 
makes it a perfect candidate for evaluation and reflection.
Moreover, we argue that similar lessons can be drawn from existing alternatives~\cite{agarwal2016developing,cronk2021quantitative,de2016priam,oetzel2014systematic,sion2019privacy,wagner2018privacy}.
At the same time, our goal is not to dismiss privacy risk assessments as an instrument;
rather,  we aim to identify and be more systematic about the challenges 
and sites of discretion that one may consider for \glspl{pra} to become an effective tool of regulation.

% relatedWork.tex

\section{Background, preliminaries and related work}
\label{s:rw}

While the terms \emph{privacy risk assessment} and \emph{privacy impact assessment} 
are often conflated and used interchangeably in the literature~\cite{wright2012privacy},
in this paper we adopt the following convention. 
We use the term \acrfull{pra} to refer to the explicit evaluation of privacy \emph{risk},
understood as the product of two components:
the \emph{likelihood} and \emph{impact} of those privacy harms.\footnote{
We note however that,  as several authors have pointed out, 
both within the expert risk assessment community and outside of, 
terminology is wildly inconsistent, 
with authors variously referring to risk as either chance,  probability or possibility;
or hazard, threat or danger, among other terms~\cite{rausand2013risk,wright2012privacy}.
}
We use the term \acrfull{pia} to refer to a broader category of assessments that
may or may not rely on a explicit \gls{pra}. Hence, \glspl{pia} include both \gls{pra} and more 
casual, informal assessments that do not formally identify \emph{risk}.\footnote{
E.g.~we would refer to a document that lists potential privacy harms a system may cause
without a explicit determination of the likelihood and impact of those harms as a \gls{pia}, 
not a~\gls{pra}.Indeed, various authors have noted that risk assessment lies at the heart of any impact assessment, 
particularly as the former conveys the notion of a more systematic and objective approach~\cite{bennett2020revisiting,raab2004future,wright2012privacy}. 
Moreover, the GDPR refers to a \gls{dpia} which, strictly speaking,  is not a \gls{pia}. 
However,  a \gls{dpia} must consider \emph{the rights and freedoms of data subjects},
among which the \emph{right to privacy}.  Hence, we consider that our analysis extends to 
\glspl{dpia} as envisioned in the GDPR. 
}
Our focus in this paper is on \gls{pra}, but we begin by tracing some of the earlier literature on~\glspl{pia}.

% Strands of research to reference:
% 1.  PIAs 
% -- including research on risk and collaborative governance 
% -- to review conditions on PIAs to suceed
Clarke traced the early origins and development of \glspl{pia}, 
finding that \glspl{pia} increasingly gained popularity since the mid-1990s
as a response to the disillusionment of decades of pro-business \glspl{fipp} and OECD guidelines
and the weak privacy legislation that they had spurred~\cite{clarke2009privacy}. 
Clarke advances two possible explanations for the emergence of \glspl{pia}:
%1
increasing pressure from the public, disappointed with privacy protections at the time, 
%2
and a \emph{natural development} of \emph{rational} management techniques, 
by adopting privacy as an additional component in risk management frameworks. 
Clarke sees this emergence in \emph{``the context of larger trends in advanced industrial societies 
to manage risk and to impose the burden of proof for [harmlessness] on its promoters''}~\cite{clarke2009privacy}.
This notion reverberates in more recent work that has interpreted the EU's regulator adoption of risk assessments, 
standardization and certification 
as a shift from command-and-control to what various authors have referred to 
as co-regulation, meta-regulation,  binary governance or collaborative governance~\cite{bennett2020revisiting,binns2017data,kamara2017co, kaminski2018binary,kaminski2020algorithmic}.
In essence, this approach espouses a philosophy of \emph{``monitored self-regulation''}~\cite{kaminski2020algorithmic}.

\glspl{pia} are said to encourage organizations to consider privacy from the inception of their projects
(in line with privacy-by-design principles) and inspire a sense of responsibility, 
enabling in turn more fully informed decision-making processes and cost-effective solutions~\cite{wright2012privacy}.
Privacy scholars have long theorized about the reasons why and conditions under which \glspl{pia}, 
as an instrument of monitored self-regulation,  would be successful~\cite{bennett2020revisiting}.
Among other factors, \glspl{pia} should engage relevant stakeholders, 
conducted in a systematic way,  be made publicly available
and subjected to audits by the regulator.
Failure to abide by these principles could render them meaningless,
enabling organizations to simply revert to self-regulation~\cite{binns2017data}.

The lack of formalism and rigorousness has haunted privacy scholarship for decades now.
Clarke observed that, at the time, many \gls{pia} guidelines were \emph{``checklist-like''}
and would not offer much beyond a compliance check~\cite{clarke2009privacy}. 
Raab aimed to \emph{``take the practice and theory of privacy protection 
beyond the stage of the merely casual use of the term `risk'},  namely, 
\emph{``a more nuanced understanding''} that would enable us to 
\emph{``estimate different degrees of privacy risk''} beyond vague statements 
such as \emph{``cookies pose a threat to privacy''}~\cite{raab2004future}.
Wright and De Hert observed that existing guidelines gave \emph{``considerable discretion to organizations}
and were \emph{``perfunctory at best, as short as two pages''},
while Bayley and Bennett found \emph{``meaningless statements that risks were identified and appropriate mitigations implemented or planned, with no hint to what those might be''}~\cite{bayley2012privacy,wright2012introduction}.\footnote{
A concern for the lack of ``formal methodologies'' has similarly crept in the context of AI-regulation~\cite{mantelero2022human}.
}
The lack of a systematic methodology further hampers the \emph{auditability} requirement.
Rather than having an established methodology or canon to evaluate a \gls{pia}, 
auditors face a plethora of unruly ad-hoc criteria and interpretations 
that make their jobs significantly harder. 

%%%												  %%%
%%% Bridge to risk analysis literature %%%
%%%												  %%%
And yet, to this day,  privacy has remained largely ignored in scholarship
devoted to the development of risk models. 
No formal methods exist to model and calculate privacy risks, 
i.e.~to understand not only what needs to be measured,  but also \emph{how}. 
Whereas a rich scholarship exists to evaluate health, environmental and technological risks,
with well-established disciplines such as toxicology,  epidemiology or exposure assessment~\cite{kammen2018should,rausand2013risk}, 
no such models have been developed for privacy. 
\gls{pia} guidelines remain vague, ad-hoc and seemingly arbitrary,
reflecting the disciplinary and political biases of their authors and institutions~\cite{agarwal2016developing,cronk2021quantitative,de2016priam,oetzel2014systematic,sion2019privacy,wagner2018privacy}.

%% Explain why privacy may be hard measure:
% 
Part of what may explain the dearth of privacy risk modeling is privacy's conceptual complexity, 
multifaceted-nature and the elusiveness of its harms~\cite{citron2022privacy,solove2006taxonomy,solove2017risk}. 
Scholars have debated the meaning of privacy for decades, 
linking privacy to theories of access, control or appropriate flow of information 
as well as its relationship to other values~\cite{nissenbaum2009privacy}.
This lack of consensus over a formal, universal definition of privacy
has led others, most notably Solove,  to opt for focusing on identifying privacy-related harms, 
regardless of whether those harms would fit under a single theory of privacy~\cite{calo2011boundaries,solove2006taxonomy}.

On the other hand, much of the empirical research on privacy has focused 
on understanding users' privacy attitudes and behavior~\cite{acquisti2015privacy,cranor2000beyond,feri2016disclosure,tucker2012economics}.
An increasingly rich literature on the behavioral economics of privacy has offered empirical evidence for 
factors that may explain the failure of notice-and-consent as a regulatory mechanism for privacy:
end-users are often unable to estimate privacy risks because of incomplete and asymmetric information, 
present-bias and the ease with which the selective disclosure of information may skew their perception of privacy risks~\cite{acquisti2015privacy}.

At the same time, there has been ample theoretical work on economic theories of privacy, 
yet again very little in terms of empirical research on privacy harms. 
Acquisti et al.\ trace three \emph{waves} of (theoretical) economic analysis of privacy, 
from its origins in the 70s and early 80s and the Chicago School of economics and their general economic arguments 
informed by market fundamentalism,
to more recent work providing formal economic models and empirical analyses. 
And yet,  as Acquisti et al.\ note,  \emph{``the economics of privacy focuses on measurable, or at least assessable, privacy components. 
Some (perhaps, many) of the consequences of privacy protection, and its erosion, go beyond the economic dimensions--for instance, the intrinsic value of privacy as a human right, or individuals’ innate desires for privacy regardless of the associated economic benefits or lack thereof''}.\footnote{
Acquisti et al.\ define the economics of privacy as hitherto develop in scholarship as 
\emph{``the analysis of the relationships between personal data and dynamic pricing''}.
Most of the work reviewed by the authors centers around how knowing information about consumers affects pricing and the distribution of consumer surplus.  
In this body of work, consumer surplus is the ultimate measure of \emph{welfare}, omitting other, less tangible social values. 
}
In other words, there remains little to no empirical evidence on the more elusive harms that result from privacy violations, 
such as loss of autonomy, trust and dignity.  
Even in what is one of the most decried privacy-infringing practices, behavioral advertising, 
there is little conclusive evidence about the privacy harms---or even benefits, 
for that matter---it causes~\cite{acquisti2020secrets}.
Moreover,  as Swire argues, \emph{the efficiency analysis of standard economic theory is impoverished compared to the concerns considered by other approaches. Economists often do not recognize an individual’s right to privacy of personal information.  Even where they do, the mere violation of a right, in the eyes of the economist, does not generally change the utilitarian calculus.  \emph{[...]} The efficiency analysis leaves out much of what people actually fear in the area of privacy protection}~\cite{swire2003efficient}.
Acquisti et al.\ refer to this as \emph{``economic `dark matter': We know it is there, but cannot directly quantify it''}~\cite{acquisti2020secrets}.

At the same time, a rich literature on risk analysis in other domains has revealed 
the limitations and politics of risk assessment as a policy and decision-making tool. 
Fischhoff et al. \ provide an illustrative overview of the limitations and shortcomings 
of addressing societal problems through the lens of \emph{risk}~\cite{fischhoff1984defining}.
They remind us that the notion of risk is inherently controversial and political.  
Riskiness depends on the underlying definition of what we are deeming risky. 
Notions of objectivity are problematic: it assumes a consensus 
or a \emph{rational} approach towards estimating risk.  It also occludes politics, e.g. ~when prioritizing individual as opposed to societal risk, 
time-effects, scope and many other factors. 
Other authors explore the contentious nature 
of settling trade-offs between risks and benefits~\cite{fischhoff2015realities,morgan1990probing}.

Perhaps closer to privacy is the field of \emph{social impact assessments}. 
Chisholm and Jesus observe that these assessments  in the context of cultural heritage are qualitative and subjective, 
there is no objectivity, and there is an implicit assumption that whoever performs the assessment 
does so in good faith~\cite{chisholm2017cultural}.
Similarly,  Glasson observes that when assessing socio-economic impacts, there is often no data, no models, 
when, in fact \emph{``all prediction methods should be some clarification of the cause-effect relationships between variables involved''}~\cite{glasson2017socio}.
Both Fischhoff and Vanclay remind us that risk creates anxiety before harm actually materializes, 
imposing a burden on people to manage risk before they materialize~\cite{fischhoff2015realities,vanclay2020reflections}.

\section{NIST's PRAM}

%NIST's PEP mission: to support the development of trustworthy information systems by applying measurement science and system engineering principles to the creation of frameworks, risk models, guidance, tools, and standards that protect privacy and, by extension, civil liberties.
Part of its cybersecurity division, NIST's \Acrlong{pep}['s] (\acrshort{pep}) mission is to 
\emph{``support the development of trustworthy information systems
by applying measurement science and system engineering principles to \emph{[...]} protect privacy''}.\footnote{
NIST \Gls{pep}, \url{https://www.nist.gov/itl/applied-cybersecurity/privacy-engineering}
}
Within this program, NIST has made available several resources, among which three main guidelines:
% NIST Privacy Framework
First, a Privacy Framework,  which describes at high-level the sort of organizational processes
and structures that organizations deploy to integrate a privacy risk assessment within their risk management strategy.
The Framework heavily borrows from NIST's Cybersecurity Framework~\cite{shen2014nist}
and it does not provide a definition of privacy or risk models for privacy harms;
its focus is entirely on organizational measures.\footnote{
NIST distinguishes between privacy risk \emph{management}, 
which relates to the processes that enable an organization to \emph{manage} privacy,
i.e. ~\emph{``the connection between business or mission drivers, organizational roles and responsibilities, and privacy protection activities''}
to \emph{``help organizations build better privacy foundations by bringing privacy risk into parity with their broader enterprise risk portfolio''},
and privacy risk \emph{assessment} as the subprocess of privacy risk management that involves the actual estimation of risk.
}
% NIST Internal Report (NISTIR) 8062
Second,  an Internal Report (NISTIR) 8062,  that introduces the concepts of privacy engineering and risk management. 
NISTIR 8062 addresses both the differences and overlap between information security and privacy, 
introduces the notion of risk as the product of likelihood and impact of adverse privacy effects 
and emphasizes that \emph{``measurability matters, so that agencies can demonstrate
the effectiveness of privacy controls in addressing identified privacy risks"}.
% Privacy Risk Assessment Methodology (PRAM)
Thirdly, and our focus in this paper, 
a \gls{pram},  which instantiates the risk management approach introduced in NISTIR 8062
by providing set of concrete steps that an organization may follow to perform a risk assessment. 

\Gls{pram} is comprised of 4 \emph{worksheets} each proposing several \emph{tasks}, 
as well as a catalog of \emph{problematic data actions and problems}. 
In the following,  we examine each of these tasks in order, identifying along the way
the sites of discretion that illustrate the challenges that \glspl{pra} pose for auditors and regulators.

\subsubsection*{WS1: Framing organizational objectives and privacy governance}
\addcontentsline{toc}{subsection}{Worksheet 1}

Worksheet 1 (WS1) addresses the elicitation of privacy requirements. 
In software engineering, requirements elicitation is the first stage in the design process, 
where the designer determines what the system must do (functional requirements) 
and which constraints may be put in the design space. (non-functional requirements, among which privacy).~\footnote{
Functional requirements define \emph{what} the system must do, 
whereas non-functional requirements, among which legal and privacy requirements, 
constrain \emph{how} one must design the system to fulfill the functional requirements. 
}

WS1 requires the analyst to \emph{``frame organizational objectives''}
such as \emph{``mission/business''} needs,  
the system's functional needs and the \emph{``privacy-preserving goals''}
that the organization may \emph{``plan to highlight or market to users or customers''}.
Next, \gls{pram} suggests to \emph{``identif[y] privacy-related legal obligations ''} as well as 
%and commitments to principles or other organizational policies''}, as this would enable the organization to \emph{``better assess the impact of data processing on [the organization's] priorities, risk tolerances, and values for individuals' privacy''}.
\emph{``any privacy goals[...] explicit or implicit in the organization's vision and/or mission''} 
and the \emph{``organization's risk tolerance with respect to privacy''}.
\begin{comment}
This includes identifying 
% 1. legal environment
\emph{``any privacy-related statutory, regulatory, contractual''} obligations;
% 2.
\emph{``any privacy-related principles or other commitments to which the organization adheres (e.g., Fair Information Practice Principles, Privacy by Design principles, ethics principles)''}; 
% 3.
\emph{``any privacy goals that are explicit or implicit in the organization’s vision and/or mission''}; 
% 4.
\emph{``any privacy-related policies or statements within the organization, or business unit''}; 
% 5.
as well as the \emph{``organization's risk tolerance with respect to privacy from \emph{[the]} organization’s enterprise risk management strategy''}.
\end{comment}

Of relevance in WS1,  in line with the conceptual foundation set in NIST's Privacy Framework and NISTIR 8062, 
is that NIST avoids committing to a particular definition of privacy.\footnote{
NIST's \emph{Privacy Framework} highlights the \emph{``broad and shifting nature of privacy''}, 
stating that privacy is an \emph{``all-encompassing concept that helps safeguard important values such as human autonomy and dignity''}
while cautioning that \emph{``human autonomy and dignity are not fixed, quantifiable constructs''}~\cite{nistPF}. 
NIST points instead to \emph{"publications that provide an in-depth treatment''}, 
citing both Solove’s and Selinger and Hartzog’s works as two sources of normative guidance~\cite{nistir8062,selinger2017obscurity,solove2006taxonomy}. 
}
While this choice reveals NIST's strategy to remain flexible, and avoid alienating organizations 
to get them to incorporate privacy into their risk management processes, 
the refusal is essentially political and normative in itself, 
for beyond the minimum legal requirements
it portrays privacy as a marketable commodity,
reinforcing the logic of self-regulation and notice-and-choice.\footnote{
As largely explored in the privacy literature, notice-and-choice has largely failed at tackling current privacy issues
as it pushes all responsibility and burden to end-users themselves,
who face a largely \emph{adversarial} market landscape, devoid of privacy-preserving alternatives~\cite{barocas2009notice,sloan2014beyond, susser2019notice}.
}
At the same time, implicit in this approach is an assumption of universality in \gls{pram} as a methodology, 
i.e.~\gls{pram} is a sound methodology regardless of the definition of privacy one considers.

\subsubsection*{WS2: Assessing system design}
\addcontentsline{toc}{subsection}{Worksheet 2}
\label{ws2}

% WS2 'Intro' sheet: 
% Purpose of the worksheet is to identify and catalog the inputs for this risk analysis.
% These inputs are the data processing operations or capabilities (i.e. data actions), the data being processed or individuals' interactions with the system/product/service, and relevant contextual factors.

Worksheet 2 (WS2) focuses on threat modeling
instructing analysts to produce a system model or data diagram flow of the system 
(Task 2 -- \emph{Supporting data map})
and project the privacy requirements elicited in WS1 onto it.
This requires the operationalization privacy goals,  privacy-related policies, principles and requirements in WS1
to concrete constraints on data operations within the system, 
e.g.~defining which parties must have access to which types of data. 
Task 3 (\emph{Contextual factors}) encourages the analyst to consider \emph{contextual factors}
such as the nature of the organizations engaged in the system (public, private or regulated industry),
or the public's perception and privacy expectations of these organizations. 
Finally, Task 4 (\emph{Data action analysis}) requires the analyst to fill a table 
to list \emph{data actions} in the system,  the data involved in such data actions, 
relevant contextual factors (as per Task 3) and a \emph{summary of issues}
which is meant to document any relevant observations related to privacy. 

All in all, WS2 closely resembles guidance commonly found in existing threat modeling methodologies.
As such,  none of these tasks are inherently flawed. 
It is their application and interpretation in the context of a privacy risk assessment
that makes them problematic.  Let us consider two particular issues. 
% The definition of the system 
Task~2 requires the production of a data diagram flow, 
a common practice in software engineering and threat modeling, 
which implicitly defines the scope of the system under consideration. 
Hence, what the organization considers to be the `system' 
will have important consequences for risk assessment.
An intuitive, obvious definition of `system' would comprise
the processes that the organization directly controls or interacts with, 
and leave as out of scope the processes and `systems' of other organizations. 
After all, an organization may have little knowledge and control over other organizations'
operational processes---even if commercial relationships may confer certain leverage to 
prime or discourage certain practices. 
However, to the extent that an organization's (data) operations are embedded within 
broader, more complex systems,  taking such an approach may lead to a myopic 
view of what is at stake,  in turn underestimating the role that the organization's \emph{subsystem}
plays within the greater system as a whole. 

% The system is a component of a broader system of interest. 
% Here the point is to identify what the system is, where the ramifications lead. 
% 
Online advertising is a case in point. 
Media publishers that wish to monetize their audiences
may ponder over whether to serve \emph{contextual} or \emph{behavioral advertising}.
Let us consider, informally, the privacy risk of installing a third-party cookie to enable advertising networks
to track readers across the web.
From the point of view of an online news organization, 
estimation of privacy risk will be wildly different depending on what 
the organization considers as \emph{system}.
The organization could consider as `system' its own websites and the content its readers see, 
i.e.~the organization's system proper. 
It may thus compute the risk of collecting and providing information about a user's visits to its websites, 
which could be innocuous enough, leading to no obvious harms by themselves, in isolation.
However,  the organization's system proper only plays a small role in the advertising ecosystem as a whole. 
To truly evaluate the impact of behavioral advertising on users, 
the organization would have to consider the vast network of cookies and other tracking mechanisms 
across the web,  and ultimately evaluate the impact of targeted advertising on society as a whole. 
An organization may also exert some control over the kind of ads it may display to its readers. 
However, the cookies it places on its websites have an impact on the ads web users see in another publisher's website. 
Hence, constraining a system's definition to the operations within an organization 
may not be truly representative of the privacy risks to which an organization contributes. 

At the same time, disentangling the contribution of a single publisher to the whole targeted ads ecosystem is far from trivial. 
Much of the value that advertisers derive from tracking users stems from the relationality of data, 
i.e.~what people that visit a particular set of websites are interested in~\cite{viljoen2021relational}.
The tracking data points from any single website, by themselves,  would be of little value. 
Similarly,  these data points, by themselves, may not lead to any obvious harms, 
in part because a single point of data may have little predictive value: 
it is the aggregation of points across websites and users, 
the emergence of patterns, that provide utility to advertisers
and represent a privacy threat to users.

Similarly, Tasks 1 and 2's focus on granular data operations may also miss the forest for the trees. 
A system model or data diagram flow is important to understand how the system works 
and how it may need to be redesigned.\footnote{
In conversations the author had with NIST representatives, 
this was one of the reported outcomes from pilot projects.
In one instance, as reported by NIST,  a CEO would deny the company 
was using a particular type of data, to which engineers would reply, 
armed with a data diagram flow, that they did process those data, 
thus triggering a more informed discussion around their operations. 
}
However,  breaking a system into atomic data operations 
and their individualized risk scores may not necessarily capture 
the combined risk of considering these data operations as a whole. 
Behavioral advertising as described above is one example of this forest-for-the-trees problem.
Another classic, illuminating example comes from the field of database privacy. 
In a groundbreaking paper,  Sweeney showed that whereas 
knowing each date of birth,  gender or ZIP-code alone would rarely lead to the re-identification 
of any individual within a database, the combination of all three pieces of information 
would enable to uniquely identify up to 87\% of individuals~\cite{sweeney2002k}. 
This illustrates how,  whereas the risk of revealing any one single attribute may be low (very low), 
the combination may be high and not equivalent to the sum of each of these attributes separately.  

% pramws3.tex

\subsubsection*{WS3: Prioritizing risk.}
\addcontentsline{toc}{subsection}{Worksheet 3}
\label{ws3}

Worksheet 3 (WS3), \emph{``Prioritizing risk''}, represents the core of the privacy risk assessment,
providing guidelines on how to estimate the two components of risk: likelihood and impact. 
WS3 defines four tasks: (1) assess likelihood, (2) assess impact, (3) calculate risk, and (4) prioritize risk, 
which we review below.

\paragraph{Likelihood. }

%Task 1 (\emph{``Assess Likelihood''}) involves the assessment of likelihood and provides key definitions to that effect. 
\gls{pram} defines likelihood as the \emph{``probability that a data action will become problematic for representative or typical individuals whose data is being processed or is interacting with the system/product/service''}.
This definition raises three main~questions. 

Firstly, what it means for data actions to \emph{``become problematic''}. 
\gls{pram} provides a \emph{Catalog of Problematic Data Actions and Problems}. 
Problematic actions include appropriation, re-identification, surveillance or unanticipated revelation, among others. 
According to \gls{pram}, these actions become problematic when they lead to \emph{problems} such as 
dignity loss, discrimination, economic loss, loss of self-determination (which includes autonomy, liberty and physical harm)
as well as loss of trust. 
Thus \gls{pram} offers to bypass the contentious definition of privacy to focus on harms that derive from privacy loss. 
In other words,  there is no specific \emph{privacy violation} that an analyst is meant to identify;
rather, the focus is on these \emph{problems} or harms,  widely recognized as byproducts of privacy loss.\footnote{
NIST representatives confirmed this was the case in conversations with one of the authors. 
Mentions of \emph{privacy threats} or \emph{privacy harms} were deemed too ambiguous and vague
in NIST's consultations with stakeholders.  Hence, NIST saw benefit in avoiding privacy as a concept, 
choosing instead to favor a panoply of privacy-related values.
}
This strategy, namely,  conceptualizing privacy through a taxonomy of harms rather through a formal definition, 
is not new, and has been previously adopted in privacy scholarship, 
most prominently by Solove~\cite{hartzog2021privacy,solove2006taxonomy}. 
However, such a strategy only seems to kick the can down the road, 
opening new challenges for the privacy analyst.\footnote{
Critiques of the \emph{taxonomization} approach point out that
by grounding privacy problems in social recognition rather than a formal definition, 
the taxonomical approach tends to identify harms as privacy-related
insofar as social commentary identifies them as so, 
arbitrarily excluding harms that may later become accepted as privacy-related, 
while failing to account for the tensions among harms that fall under the privacy umbrella~\cite{calo2011boundaries}. 
}

While it is hard to contend with the argument that privacy is a slippery and hard to pin down concept, 
values such as dignity or autonomy are equally slippery and open to interpretation,
%There is no universal consensus over what these values mean either, 
thus representing yet another point for an organization to exercise discretion,
i.e.~encoding their own, self-interested version of privacy by cherry-picking how they 
decide to incorporate these proxy values into the analysis, 
e.g.~giving them more or less weight, or simply ignoring~them. 

% Here I'm raising the problem of whether some problems may even register for the analyst. 
% It's clear that not every problem is tractable -- the space of problems is just too big
% so we're focusing 

%Lack of actuarial models + intractable space of problems --> simplification leads to further points of discretion
Moreover,  the space of potential problems is far too big for an analyst to consider in its totality, 
a problem compounded by the nonexistence of actuarial models for privacy.
One problematic action may lead to a wide range of consequences,  
from the most banal to the catastrophic, and estimating their probability depends 
on a large number of factors that is not easy to model or anticipate. 
Consider the AOL search log anonymization blunder, where two journalists at The New York Times managed to re-identify 
several searchers~\cite{barbaro2006face}.  While a (better) privacy risk analysis may have anticipated the risk of re-identification,  
the consequences and their likelihood would have been far harder to determine.\footnote{
Even if the journalists followed ethical disclosure practices---as they did, 
requesting permission from re-identified searchers to publish their results---
other parties may not have been so careful. }
Privacy harms are notoriously elusive. 
The loss of trust in search engines as a result of loose data sharing practices cannot be easily quantified.
How individuals in the AOL dataset may have approached online search from that moment on, 
or how other individuals not included in the dataset,  whether using AOL or another search engine,  
may have become mistrustful of search engines has an impact on society overall that, 
to this day, we do not know how to measure. 
The expectation that one's searches may be exposed or monitored may restrain people's self-expression,  
leading to self-censorship or anxiety over ``risky~searches''. 

Citron and Solove have identified numerous factors that illustrate the elusiveness of privacy harms, 
explaining why US courts have consistently failed to acknowledge them~\cite{citron2022privacy, solove2017risk}. 
Harms may be small but numerous, often repeated over time and perpetrated by multiple entities.
Or they may be tiny but affect a very large number of people.
While any of these harms may be small enough to dismiss and even fail to register 
in a risk analysis,  the combined effect could be significant and harder to ignore. 
In many cases, the harm is not even knowable, 
the loss of trust in search engines as described above a case in point, 
but also the chains of transmission that go beyond the initial disclosure, 
e.g.~an organization A that discloses data to organization B
may not be able to account for B's subsequent uses of data.
% Subsequent reuses and risk assessments
%An organization A may choose to disclose data to some other organization B 
%given certain assumptions about what B may do with the data. 
%However,  B's subsequent behavior may deviate from A risk analysi s could legitimize it...
%to further expose and reuse the data in ways that 
%A would not be able to imagine or account for, and vice versa. 
Harms are also easy to overlook due to the relational and contextual value of data. 
Seemingly innocuous pieces of information may become sensitive in a different context, 
or when combined with additional data~\cite{narayanan2008robust,nissenbaum2009privacy}. 
Finding causality between privacy breaches and privacy harms is also challenging,
because misuses are not widely known or advertised, 
i.e. ~malicious entities have incentives to keep their practices secret, e.g.~which data they are using, 
what they are combining it with.
More generally,  it was the realization that it is impossible to provide any absolute guarantees 
of privacy risk or protection that motivated the push for differential privacy~\cite{dwork2010difficulties}. 
This should indeed give us pause when attempting to measure the likelihood of a privacy harm.

In short,  a privacy analyst cannot be realistically expected to account for every possible situation and harm, 
let alone be able to measure their \emph{impact} as Task 2 below requires. 
This incommensurable complexity forces analysts to engage in two types of simplification.
First,  the analyst must select a strategy to scope the problem space within which problems are considered at all, 
leaving other problems out. 
Second,  the analyst must select a threshold beyond which certain events become \emph{problematic}.
Both strategies lead to bias in the selection of which types of problems make it into the risk assessment, 
further reinforcing the idea that organizations can mold risks assessments to fit their own agenda, 
rather than aiming for socially beneficial outcomes. 
This is how risk assessments may become arbitrary and biased, 
in spite of the veneer of legitimacy that (pseudo-)statistic methods may confer to it.

% about who the users / non-users are
Secondly,  who the \emph{representative} or \emph{typical} individual is.  
% First,  how do we do this? 
% Two approaches: 
% 		first, we model our population, understanding how they may be impacted by harm
% 		second,  we model our population based on some other factor (demographics, etc.) and measure harm based on that
We may consider two possible approaches to determine who this typical individual is. 
One,  in terms of demographics relevant to the organization's bottom-line (e.g.~in terms of age, revenue, loyalty). 
Another,  in terms of actual risk, i.e.~the pool of individuals who would typically bear the brunt of privacy harms.
Intuitively, the second approach would be the most rigorous. 
However,  this is a chicken and egg problem:  to determine who the \emph{representative} individuals are,
we need to perform an assessment for all possible representative individuals,
thereby defeating the very purpose of this simplifying strategy.
Not only the likelihood and impact of suffering a privacy problem may vary widely for different individuals, 
but also their ability to deal with them, 
as people may have vastly different abilities and resources. 
Affluent or tech-savvy people are likely better equipped to prevent or address privacy harms, 
e.g.~by their ability to hire lawyers or have measures in place that shield them from the organization's missteps.
Thus a \emph{distributional inequity} problem surfaces~\cite{wilson2001risk}.

%The inequity problem is compounded by the fact that the \emph{representative individual} framing 
%implies that there are and we may ignore non-representative or atypical individuals 
%whose data may be processed by the system, 
%or about non-users of the system who, because of the relational nature of data, may be affected by the system.
Organizations, even those conducting \glspl{pra} in good faith, are likely to prioritize 
their own users or clients,  thus introducing a bias in who they calculate risk for. 
Consider the secret collection of images of people from social media and the web to develop facial recognition technology,
as in the Clearview scandal~\cite{roussi2020resisting}.
While the most immediate privacy violation that led to widespread condemnation relates 
to those individuals whose images were scraped, 
the privacy harms for individuals who are not even in the dataset and upon whom face recognition technology 
is later deployed could be disastrous and yet more easily overlooked. \footnote{
This is yet another example of how the relational nature of data challenges
the notion of a \emph{representative} individual. 
While the social, relational and interdependent nature of privacy has been amply theorized, 
an individualistic perspective still dominates public discourse, thereby sweeping under the rug 
the kind of networked, relational harms that a more capacious notion of privacy would surface~\cite{citron2022privacy, cohen2013privacy,regan2002privacy,viljoen2021relational}.
}

Thirdly, the actual computation of likelihood.
\gls{pram} suggests to measure \emph{``the probability of occurrence for each potential problem''} on a scale from 1 to 10, 
while conceding that \emph{``organizations can use any scale they prefer as long as they use the same scale throughout the process''}. 
The elusiveness of privacy harms means that computing their probability of occurrence is a very complex task. 
The absence of actuarial models means that likelihood estimation must be done from scratch,  
with no existing references to compare and gauge, and therefore to readily challenge estimations.
Similarly, giving organizations' the freedom to choose their own measurement scale 
not only represents yet another site of discretion, it also leads to auditing challenges. 
Auditors (as well as the broader public or whoever intends to examine an organization's \gls{pra}) 
must grapple with different estimation methods and scales,
a significant auditing burden that ultimately hinders accountability.

%In short,  instructing an analyst to estimate the 
%\emph{``probability that a data action will become problematic for representative 
%or typical individuals whose data is being processed or is interacting with the system/product/service''} 
%raises a myriad of challenges for both analysts and auditors. 
%The points above seek to illustrate the vast room for discretion organizations have,
% to determine: what problems? for whom? how to measure their probability? 
%and the myriad configurations of any of these variables points to the challenge 
%of conducting a privacy risk assessment in a principled, rigorous way.

\paragraph{Impact.}
Task 2 (\emph{``Assess impact''}) gives organizations even more discretion.
In recognizing that \emph{``it may be difficult for an organization to assess the impact of these problems''}, 
NIST suggests that \emph{``should \emph{[the analyst]} be unable to \emph{[assess direct impact on individuals]}, 
organizational impact factors as secondary costs absorbed 
by the organization can be used in lieu of or in addition to direct impact assessment''}.\footnote{
As organizational impact factors, \gls{pram} proposes \emph{non compliance costs} such as fines or litigation expenses;
\emph{direct business costs} such as loss of revenue;
\emph{reputational costs} such as brand damage or loss of consumers' trust;
and \emph{internal culture costs} such as impact on employee morale, as well as \emph{``other costs an organization wants to consider''}.}

Allowing organizations to consider organizational rather than societal impact
has been touted as a strategy to encourage them to embrace~\glspl{pia}.
After all,  it makes sense that organizations will be more willing to address privacy problems
if they realize these affect their bottom-line. 
Besides, as privacy harms are often elusive,
tying them to more tangible ``proxy'' costs could make their estimation easier. 

Yet such a rationale can easily backfire if an organization's incentives are misaligned with society's. 
%Choosing to assess an organization's absorbed costs as opposed to direct impact on individuals and society 
%has two main unintended effects. 
First,  privacy harms may simply not translate into costs for an organization;
organizations may also be able to externalize these costs and thus ignore privacy. 
Tech companies like Google or Facebook may be able to externalize most of the societal cost stemming from the loss of opportunity 
that results from (discriminative) targeted advertising, 
due to the highly lucrative duopoly they hold over the online advertising market~\cite{imana2021auditing,speicher2018potential}. 
%\eb{add more refs, particularly for monopoly stuff}
Secondly,  such an approach implicitly legitimizes risk management strategies that do not seek to remediate 
privacy risks per se but, rather, simply mitigate the impact of such risks on the organization,
e.g. ~in mitigating reputational costs, an organization may invest in PR rather than privacy;
in avoiding regulatory fines, an organization may lobby public officials to weaken privacy regulations~\cite{waldman2021industry}.
%Hence,  using organizational impact factors as a proxy for privacy harms is, to put it mildly, a double-edged sword. 
And even if an organization acts in good faith and attempts to measure the impact on individuals and society, 
similar challenges arise as discussed for the estimation of likelihood above, 
such as the incommensurability of all the potential avenues for harm
or the diversity and breadth of the affected population. 
%As discussed above, Citron and Solove have amply documented how courts have often dismissed privacy harms 
%because there is weak evidence of their impact~\cite{citron2022privacy,solove2017risk}.
%This may enable both good faith and bad faith organizations to dismiss many privacy harms. 
%Fiven the glaring dearth of empirical evidence on privacy harms and their impact, 
%neither privacy risk analyst nor auditor have previous actuarial models on privacy harms on which to base their analysis. 
%Their analysis must start from scratch, i.e.~from the very collection of empirical evidence of privacy~harms. 
% there is no evidence of harm -- again, think about actuarial models -- low probability,  low evidence of harms, intangible harms 
%-- this is the perfect recipe, trojan horse for organizations to do whatever they want

The empirical estimation of privacy harms raises additional conundrums, 
as it must to some extent involve experimentation, 
potentially subjecting people to privacy harms we wish to identify and model.  
To the extent that these experiments are run under controlled conditions and strict 
ethical supervision, it may be possible to avert or minimize some of these harms. 
Cases such as Instagram's internal study of teens' mental health, 
or Facebook's experiments on emotional contagion, however, 
show that those organizations that have the data and manpower to perform such experiments
may not always do so with sufficiently robust guardrails in place~\cite{bird2016exploring,verma2014editorial,wells2021facebook} 

Another conundrum relates to the measure of privacy impact. 
Economic analyses as reviewed in Sect.~\ref{s:rw} would likely resort to dollars as a measure of consumer welfare.
Other risk measures such as fatal accident rate, lost-time injury or reduction in life expectancy make sense in contexts such as health or occupational safety~\cite{rausand2013risk}.%  \eb{more refs, examples?}
However, such measures cannot capture the wide range of harms that privacy encompasses. 
As Acquisti et al.\ argue, \emph{``costs of privacy across scenarios are arguably impossible to combine into a definitive, 
aggregate estimation of the `loss' caused by a generalized lack of privacy''}~\cite{acquisti2020secrets}.

PRAM bypasses this problem by suggesting that the analyst determines
\emph{``on a scale from 1-10 the estimated effect of each potential problem''}. 
Those values are then \emph{``added to calculate organizational impact per potential problem''}.
%Yet, as we have argued,  there are no actuarial models 
%to determine the impact of each of the problems in the NIST catalog. 
%Hence, analysts are on their own when it comes to assigning a value from 1 to 10. 
%In other words, the analyst may compute an individual impact score for \emph{non-compliance},  \emph{business} and \emph{reputation} scores, 
%then add these up to obtain \emph{total impact}. 
Hence, \gls{pram} assumes a linear cumulative relationship between individual problems. 
There is however no evidence of a linear relationship between these values, 
i.e.,~a loss of trust of magnitude 7 plus a loss of autonomy of magnitude 5 
would result in an impact of magnitude 12.\footnote{
Moreover,  such operations may not even make sense, 
mathematically, when using an ad-hoc scale.}
Such losses may discount or reinforce each other, 
leading to over and underestimation of aggregate effects. 
Assuming linearity also prompts questions about what the resulting impact represents, 
since we are combining different types of losses, 
i.e. what does a combined impact of magnitude 12 represent? 
Similarly, it also assumes linearity of the scale across different types of privacy problems, 
i.e. it assumes a dignity loss of value 5 is equally severe as a liberty loss of value 5.
To the best of our knowledge, no evidence for any of these assumptions currently exists.
These issues propagate on to the next task \emph{``Calculate risk''}.

\paragraph{Risk.}

Task 3 (\emph{``Calculate risk''}) simply prompts the analyst to multiply likelihood and impact 
and add the resulting risk scores up per problematic data action.
Our observations mirror those we have made above about linearity and composability.
In short,  \gls{pram} computes a total score operating on individual likelihoods 
and impact scores in a way that is both conceptually and mathematically incorrect.  
%For details, we refer the reader to Appendix~\ref{apdx:ws3}.

\paragraph{Management.}

Lastly, Task 4 (\emph{``Prioritize Risk''}) provides suggestions for the organization to prioritize risks, 
as part of a risk management strategy.
\gls{pram} suggests that an organization may choose \emph{``prioritization methods that provide 
the best communication tool for their organization and that best support decision-making''}, 
once again legitimizing methods that may align with the organization's bottom-line as opposed to societal good. 
At the same time,  \gls{pram} proposes two sample prioritization methods, 
both roughly implying that those actions with the highest risk score be dealt with first. 
%The first is a simple percentage ranking method that 
%amounts to prioritizing the most risky data actions above others. 
%The second prioritization method (\emph{``Top 5 outliers''}) involves mapping coordinates of risk and impact 
%to identify the problems that have the greater combined likelihood and impact values. 
%This strategy prioritizes problems with simultaneously high likelihood and impact values, 
%as opposed to the first strategy, where actions with a combined high risk,  even if either they may be relatively low impact or low likelihood, 
%would get higher priority if their risk score is among the highest. 

While prioritizing the riskiest operations makes sense---even if, at this point, 
we hope to have shown how the accumulation of conceptual errors and indeterminacies 
may render these values virtually meaningless or,  worse still,  the result of 
a biased analysis that seeks to legitimize the organization's operations---
it is far from trivial that this is the right prioritization strategy
or that it should 
%Let us however assume for the sake of analysis that,  at this point, 
%the risks scores obtained are somewhat reasonable and meaningful. 
%A prioritization strategy is far from trivial, and should rarely 
simply boil down to simple math ordering or mapping, as \gls{pram} suggests. 

There are a number of elements that \gls{pram} does not even surface. 
Firstly,  the quintessential question: how much risk is too much? 
\Gls{pram} offers no guidance to determine whether 
an operation is simply too risky and not worth pursuing at all. 
There is no mention of balancing risks and benefits, 
and how to perform such balancing. 
Ideally, we would expect that risky operations should only be pursued
if they can be justified for the greater good and, even then, 
minimizing the risks to the extent that is viably possible. 
\gls{pram} does not address the estimation of benefit, 
or the need to perform a benefit-risk-cost analysis~\cite{fischhoff2015realities}. 
Some operations may entail risks which are too expensive to manage, 
other operations' benefits may be too slim. 
Both should dissuade an organization from carrying such operations.
Equity issues resurface when we consider the distribution of risks of benefits. 
The benefits may be great,  yet for a selected few at the expense of the majority, who face increased risks.
Benefits may also be immediately enjoyable, as opposed to risks expected in the future, 
rendering the latter easy to discount. \footnote{A classic example is climate risk, 
where the benefits of extracting oil can be immediately enjoyed, 
while future risks,  even if already apparent at present, are easily discounted.}
\gls{pram} bypasses this complexity by suggesting 
that organizations should manage these risks at their own discretion.
%In their report on privacy engineering and risk management, 
%NIST warns that \emph{risk can be managed, but it cannot be eliminated}~\cite{nistir8062}.
The risks of trusting organizations with this kind of assessment are well known. 
As reported by the WSJ, Facebook would argue that the risks of using Instagram 
\emph{``are manageable and can be outweighed by the app's utility''}, 
while a mounting body of independent research may call this assessment into question~\cite{wells2021facebook}.  
%\emph{``Facebook’s research indicates Instagram's effects aren't harmful for all users.  For most teenagers, the effects of `negative social comparison' are manageable and can be outweighed by the app’s utility as a fun way for users to express themselves and connect with friends, the research says.''}
%\url{https://www.wsj.com/articles/facebook-knows-instagram-is-toxic-for-teen-girls-company-documents-show-11631620739}

\section{Lessons and discussion}

In this  section we summarize and discuss some of the lessons that 
we have drawn from our analysis of \gls{pram}.

\subsubsection*{Sites of discretion}

Privacy risk assessment guidelines such as \gls{pram} introduce multiple sites of discretion
that adversarial organizations---namely, those whose interests are misaligned with privacy---
can use to engage in performative compliance.
We have identified the following in \gls{pram}.

\paragraph{Defining \emph{privacy}. }

It goes without saying that to assess privacy risk we need a definition of privacy, 
or at least some notion of what it encompasses.
%Scholars agree that privacy promotes values such as autonomy, dignity and freedom, among others, 
%but there is no universally agreed upon definition. 
%
%In systems design and engineering,  perhaps even more so than in other disciplines, 
%privacy is likely to be misunderstood and misinterpreted,  spelling trouble for analysts who lack the expertise and training, 
%while bolstering those who seek to weaponize this lack of consensus to their advantage~\cite{waldman2021industry}.
A narrow definition of privacy is likely to underestimate legitimate risk,
while a biased, interested definition of privacy is likely to downplay risks that do not fit the analyst's agenda. 
The lack of consensus and standards represents a challenge for auditors,  
who must grapple with potentially unique definitions of privacy across organizations. 

Yet this is an inescapable problem.  Even under universal definitions of privacy, 
such as Contextual Integrity,  the operationalization of privacy varies across contexts, 
e.g.~information flows which are appropriate in one context may not be in another. 
As a result, there is no easy template or checklist that auditors can use to verify the validity of a \gls{pra}.
A subtle understanding of contextual factors---e.g.~informational norms and expectations, 
goals and values---is necessary to calculate risk.
This implicitly represents an endorsement of the meta-regulatory approach
insofar as it acknowledges that organizations understand the context
on which they operate better than the regulator.
It also highlights how given too much latitude, 
adversarial organizations may weaponize such epistemic asymmetry.
Hence, guidelines should steer analysts towards socially meaningful notions of privacy, 
while auditors must place extra care to scrutinize
the privacy notion that organizations operationalize in their \gls{pra}.

% Worksheet 1 - Task 1 suggests the org lists privacy goals or properties that they may want to market to consumers. 
% Here privacy is an optional, added-value-like commodity, rather than the desirable state of system design (i.e. what it ought to be). 
% It's furthering a detrimental discourse, i.e. that privacy is an optional requirement, not an expected baseline. 
% This is reinforced in Task 2, where the org is encouraged to list legal privacy requirements and any other privacy "goals" or commitments the company abides by.

\paragraph{System definition.}

% In a nutshell: in privacy, the system definition may not be the service. 
% the service may be embedded within a system of systems 
% or it may be a service to others, in the way the technology is deployed
% so it's another site of discretion. 

\gls{pram}'s WS2 illustrates how an organization's definition of service or system 
may not represent a meaningful unit of observation to assess of privacy risk. 
Understanding that the organization may only play a small role or represent 
a component of a greater system is important and likely to be overlooked. 
Because data is a non-rivalrous and non-excludable good, 
as well as the relational nature of privacy, 
constraining privacy risk to an organization's system or operations
may lead to unreliable assessments. 
As we review below,  each organization's operations within a larger system may pose very small risk in itself;
yet altogether accrue to significant risks.

\paragraph{Lack of empirical evidence,  actuarial models and cognitive biases.}

The lack of empirical evidence of privacy harms means that 
\glspl{pra} today are more of an art and practice than a science. 
They are performed based on intuition and principles rather than actuarial models. 
Because of this,  there is a large element of subjectivity in \glspl{pra} 
that make this instrument particularly vulnerable to analysts' biases, 
whether they are steered by the organization, their own personal beliefs, or both.\footnote{
Early accounts reveal how external analysts cannot be realistically expected to 
exercise full discretion in their recommendations and become subjected to their employers' pressures.
As Waters reports: \emph{``Clients will rarely welcome a recommendation that an entire project be taken back to the drawing board and fundamentally be re-designed [...] it is unrealistic to expect \gls{pia} practitioners to make such recommendations}~\cite{wright2012privacy}.
}
Consequently,  not only are they hard to audit,  
but may also fall victim to auditors' capriciousness,
as it is unlikely that they share the same subjectivity and values as the analyst. 

\paragraph{Risk accretion.}

Some risks may be deemed negligibly low on an individual basis.
By systematically failing to address them across individuals or through time, 
we incur in risk accretion that steadily and insidiously erodes privacy, 
a sort of privacy death by a thousand cuts.\footnote{
As Wilson and Crouch note, \emph{``For \emph{[most]} risk analysis \emph{[...]} a great deal of precision in calculation is not necessary: if large it is obvious that the risk exceeds the benefit and if negligible it is obvious that the benefit exceeds the risk''}~\cite{wilson2001risk}.
}

\paragraph{Linearity and composability.} 

In addition to the lack of empirical evidence, 
\gls{pram}'s WS3 illustrates how baking assumptions of linearity and composability 
leads to pseudoscientific assessments.
While risk analysis is not an exact science and often relies on 
approximate divide-and-conquer strategies to deal with complex systems,
those assumptions must follow evidence-based model approximations. 
Failing to do so leads to arbitrary, meaningless assessments. 
As Stirling observes, \emph{``under uncertainty,  attempts to assert 
a single aggregated picture of risk are neither rational nor `science based'{''}}~\cite{stirling2007risk}.

\paragraph{Distributional effects.}

As our analysis of \gls{pram}'s WS3 illustrates, 
privacy harms may not only be unevenly distributed, 
but the affected population may be unequally equipped to cope with those harms. 
Organizations are likely to prioritize their own clients or potential users.\footnote{
Acquisti et al.\ observe that \emph{``privacy trade-offs are inherently redistributive''}, as decisions affect stakeholders differently, and stakeholders' interests \emph{``are rarely aligned''}. \emph{``The theoretical goal of achieving desirable societal outcomes inevitably forces policymakers to tackle thorny questions regarding whose welfare they want to prioritize. Either by intervening with regulation, or by letting market outcomes determine levels of privacy across domains, they inescapably favor one or the other stakeholder''}~\cite{acquisti2020secrets}.
}

%power users >> users  >> potential users >> society as a whole

\paragraph{Benefit-cost-analysis.}

Guidelines like \gls{pram} do not address the trade-off between risk, cost and benefits
inherent in risk management. 
In other words,  such guidelines do not offer guidance as to 
what is too great a risk, or when benefits outweigh risks,
leaving the legitimacy question unanswered.  
This has been a long-standing issue in the field, 
with Clarke observing that \emph{``[q]uite simply, there is no vehicle for answering the question:
`Should a particular [practice] or system exist at all?'{''}}~\cite{clarke2009privacy}.
The \gls{pra} approach gives organizations free rein to settle these considerations, 
assuming that such a comparison is somehow rational---i.e.~do benefits outweigh risks
according to estimated values?---when it is, in fact,  political.

\subsubsection*{On contextual integrity and utilitarianism.}

\glspl{pra} encode a utilitarian stance to privacy,
as an organization may legitimize its operations if it 
can show that the benefits outweigh the privacy~risks.\footnote{
As we have argued earlier, given the elusiveness of privacy harms, 
and the often partial role that an organization may play in a larger privacy-infringing system, 
it is easier to showcase the benefits and downplay the risks, 
e.g.~social media platforms can trumpet the rich interactions that advertising revenue subsidizes
while easily downplaying insidious privacy harms related to hypertargeting and personalization 
---harms that we currently do not understand well in spite of worrying evidence from 
cases of political hypertargeting and emotional manipulation~\cite{verma2014editorial}.   
}

The theory of \Gls{ci} enables us to see how this may be problematic. 
According to \gls{ci},  privacy is defined by long-standing norms and expectations
about information flows in a particular context, with context understood as a social sphere with defined goals, ends and values, 
such as health care,  the workplace, or education~\cite{nissenbaum2009privacy}. 
Informational norms are supposed to have evolved to promote such ends, goals and values,
e.g. doctors ask their patients very sensitive questions 
under strict confidentiality conditions; confidentiality supports the overall goal
of promoting patients' health.  Patients are (at least in theory) 
reassured to speak freely and frankly about their symptoms,  enabling in turn better diagnosing.  

While \glspl{pra} should be informed by norms and expectations, they implicitly upend them. 
Consider doctors performing a \gls{pra} for each patient and piece of data, 
rather than abiding by the principle of confidentiality. 
Doctors may deem benefits may outweigh a negligible risk to patients, 
e.g.~an employee's doctor may disclose to a mistrustful employer, in exchange for a sum of money, 
that the employee has the flu, considering that such a disclosure would seldom cause the patient much harm.
While benefits may outweigh risks in many situations,
engaging in such ad-hoc analyses opens a Pandora's box, 
subverting long-established informational norms that support the fabric of society. 
This is not to say that a principled \gls{pra} cannot faithfully align 
with informational norms and expectations;
rather,  it is the multiple sites of discretion we have outlined above 
that threaten the legitimization of deviations from such long-established principles, 
particularly in case of new technologies, where norms may not be readily apparent~\cite{nissenbaum2009privacy}.

Ironically,  informational norms underpinning \gls{ci} are the result of a \gls{pra} of sorts.
Many norms have been established through time by trial and error, 
the result of a sort of social and historical deliberation process in pursuit of certain values.\footnote{
Confidentiality is one of the Hippocratic Oath's core principles, 
dating all the way back to ancient Greece. 
}
As these norms represent the result of a subtle process of political consolidation 
into normative systems that purportedly support socially desirable outcomes, 
they encode a trade-off between the benefits and risks of certain information flows.\footnote{
Not all norms are solidly established or have a long tradition. 
As Nissenbaum observes, some norms are in flux, 
while others may be the result of recent but profound social changes.~\cite{nissenbaum2009privacy}.}
Hence, we may argue that these norms are the result of an implicit socially-elicited \gls{pra}.\footnote{
Nissenbaum acknowledges that \gls{ci} is implicitly conservative, 
and that some norms may not support the ideal values in a context, 
resulting from asymmetrical power struggles that do not maximize social welfare~\cite{nissenbaum2009privacy}. 
}

%% Here technocratic-ness, which I should connect with privacy engineering as a whole
\subsubsection*{Privacy engineering, legitimacy and the technocratization of privacy.}

\glspl{pra} embody a technocratic approach to privacy because they rely on the premise that 
they \emph{``offer a comprehensively rigorous basis for informing decision-making''}~\cite{stirling2007risk}, 
i.e.~a \gls{pra} is assumed to be scientific,  technical and objective;
an analyst that performs a sound \gls{pra} must arrive at the correct decision and design.\footnote{
The privacy literature is littered with statements that reinforce this notion.  To mention a couple of examples:
NIST states that \emph{``[m]easurability matters, 
so that agencies can demonstrate the effectiveness of privacy controls in addressing identified privacy risks''}~\cite{nistir8062}.
Oetzel and Spierkermann's state that \emph{``because \emph{[a PIA]} offers a risk management 
approach that includes standardised procedures\emph{[,  it]} should lead to concrete technical improvements, 
[overcoming] the largely qualitative approach of the legal compliance domain''}~\cite{oetzel2014systematic}.
Gellert refers to risk as \emph{``an instrument that allows for the transformation
of uncertain future dangers into certain future dangers. 
Such transformation is possible because of the use of statistics and probabilities}.~\cite{gellert2015data}
}
To include \glspl{pra} as part of the privacy engineering toolkit,  as NIST does, 
further reinforces this idea.  
And while privacy engineers play an important role in the design of privacy-friendly systems, 
as amply documented elsewhere~\cite{gurses2015engineering}, 
portraying \glspl{pra} as a purely technical affair is both misleading and misguided.
Examining a privacy engineer's role in establishing a system's legitimacy helps us illustrate why.

In a nutshell, a privacy engineer implements data minimization strategies.
As such, privacy engineers play a key role in minimizing privacy risks,
preventing mission creep and illegitimate uses of data. 
In assessing a system's legitimacy,  let us simplify and consider two factors. 
Firstly, the functional purpose itself, i.e.~what the system is supposed to do or its \emph{functional requirements}. 
Secondly, its implementation, i.e.~how the system is implemented,  
which depends on a set of \emph{non-functional} requirements, such as security and privacy. 
On the one hand, a \emph{prima-facie} legitimate system may be deemed illegitimate 
because its implementation causes undue privacy risks or concerns, 
such as {e-mail} services where users' messages are routinely scanned for profiling 
or an instant messaging service where messages are sent unencrypted, 
so that third parties can eavesdrop users' conversations.
On the other hand,  a system may be deemed illegitimate because its very purpose is privacy-infringing, 
such as (arguably) behavioral advertising or indiscriminate surveillance.

A privacy engineer may redesign these systems to
collect or expose the bare minimum amount of information to fulfill the desired functionality, 
e.g.~implementing end-to-end encryption so that only sender and recipients 
are able to read exchanged messages. 
However, when the system itself poses a fundamental risk to privacy, 
there is nothing a privacy engineer can do to make the system legitimate. 
A case in point is Google's ongoing efforts to phase out third-party cookies, 
addressing the most egregious privacy concerns in \emph{how} behavioral advertising is implemented
to legitimize what is still arguably,  an illegitimate secondary use of user data~\cite{mccarthy2020google}.

In short, privacy engineering enables us to mitigate privacy threats insofar 
as the functionality we wish to implement does not lead to privacy risks \emph{in itself}. 
However, it is not equipped to answer whether a system is legitimate or not
because that is, in essence, a political and ethical question that lies outside the realm of engineering. \footnote{
That is not to say that engineering practice \emph{is not} political, 
as a rich literature on the politics of technology has amply documented elsewhere~\cite{winner2017artifacts}.
}
That is why trying to pass privacy risk assessments as part of privacy engineering practice 
is so insidiously perilous: it could help legitimize, 
through privacy technologies,  privacy invasive systems at heart.

% discussion.CPDP

\subsubsection*{On (collaborative) governance.}

%Sites of discretion create a nightmare for governance. 
%-- hard to audit 
%-- performative compliance 
%-- it's not simply follow the method and you'll be good -- that's not this kind of engineering
%-- the absurdity of asking the mining company to do the environmental impact assessment 

Our analysis raises numerous questions about governance.
For those organizations whose economic imperatives are fundamentally misaligned with privacy, 
\glspl{pra} offer an opportunity to engage in performative compliance. 
Under the model of collaborative governance, 
organizations are meant to conduct \glspl{pra} under light supervision, 
even if they may be potentially subjected to audits.

% Discretion: don't let companies run their own risk assessment 
% and: you should have parameters set in advance, an EPA or something!

This governance model seems to be set up to fail.
In the same way that we would not trust a mining company 
to conduct an environmental impact assessment
or a chemical plant to decide pollution levels, 
the lack of clear guidelines on what is too much risk
and what benefits may justify it pose a challenge for governance. 
While the US's \Gls{epa} sets mandatory federal standards for drinking water
and air quality,  there is no such standard or thresholds for privacy risk~\cite{lewis1985birth}.
Pushing for the adoption of \glspl{pra} would require an independent body
subjected to democratic authority (e.g.~a~\acrshort{dpa}) to set and oversee acceptable privacy risk levels. 
Organizations could then run their own assessments to determine whether or not their activities fall beyond those levels.
However with no empirical models on privacy harms, 
it is unclear that \glspl{pra} are viable and a good idea in the first place. \footnote{
Granted that they may provide a tool to diligent and good-willing 
organizations to brainstorm about the privacy implications of their operations. 
}

Collaborative governance also seeks to leverage organizations' expertise, 
relegating the regulator to a mere auditing role. 
Paradoxically, the lack empirical models means that, 
to test an organization's assessment, 
auditors would likely need to perform each \gls{pra} from scratch, 
as they have no actuarial models to use as a reference. 
As recent experience with \glspl{aia} show, 
\emph{``[it] requires a lot of energy to bridge the gap between 
getting the audit results and then translating that into accountability''}~\cite{ng2021auditing}.
% https://themarkup.org/the-breakdown/2021/02/23/can-auditing-eliminate-bias-from-algorithms
Hence,  the purported benefit of leveraging organizations' expertise seem flimsy.

\subsubsection*{Takeaways and future work}

Our analysis has hopefully shown that we should tame our expectations 
as to what we can expect from \glspl{pra}. 
Whereas as a heuristic or brainstorming exercise they may bring value,  
we should be critical and careful of how much credit we give them and how they guide our decisions. 
The potential for misuse is vast, 
while the regulator's ability to prevent such misuse is doubtful, 
in light of the lack of resources for regulatory agencies.

A \emph{``science''} of privacy does not currently exist. 
This does not mean that we should not attempt to develop it, 
try to understand privacy harms, gather empirical evidence. 
Yet to the extent that our current governance approach depends
on such models, we must find alternative guidance. 

As stated earlier, \gls{ci} relies on existing norms and expectations, 
implicitly the result of a \gls{pra} of sorts. 
\gls{ci} encodes a preference for the precautionary principle, 
honoring expectations and existing principles rather than taking an all-is-up-for-grabs approach
which, as similarly to Stirling's arguments in the context of human health and the environment,
\emph{``provides a general normative guide, to the effect that policy making under uncertainty, 
ambiguity and ignorance should give `the benefit of the doubt' to the protection of human health 
and the environment, rather than to competing organizational or economic interests''}~\cite{stirling2007risk}.
Yet the implementation of \gls{ci} is not without challanges of its own. 
Often values are in flux and norms not readily apparent for new technologies. 
Similarly,  new or updated informational norms may support contextual ends and values,
in which case an overhaul of entrenched norms would be in order. 
As Stirling argues, \emph{``if what we seek [is]
to remove the need for subjectivity, argument, deliberation and politics, 
then precaution offers no such promise.  
Instead, it points to a rich array of methods that reveal more explicitly and accountably
the intrinsically normative and contestable basis for decisions, 
and the different ways in which our knowledge is so often incomplete. 
This is as good a `rule' for decision-making, as we can reasonably get''}~\cite{stirling2007risk}.
Interestingly, \gls{ci} offers such a \emph{normative and contestable basis for decisions}.
Hence,  in the spirit of \gls{ci}, finding ways to integrate a deliberative framework around information flows
that support contextual ends and values represents an alternative worth pursuing in the future. 
Stirling points to several methods that may support such an approach
while avoiding the pitfalls of the pseudoscientific approach implicit in \gls{pra}~\cite{stirling2007risk}.

% conclusionCPDP.tex

\section{Conclusion}

In this paper we have examined \acrfull{pra} as an instrument of collaborative governance, 
taking NIST's \acrfull{pram} as a case study. 
We have shown that multiple \emph{sites of discretion} and a lack of empirical models 
on privacy harms threaten their applicability and undermine the very premises of their utility. 
\glspl{pra} are touted as technical, objective,  essentially technocratic means of governance,
occluding the ability of adversarial organizations 
to use them as instruments of \emph{performative compliance}.

We have conducted a detailed analysis of where \gls{pra} may fail or be misused, 
the challenges auditors face,  and the existence of alternative deliberative frameworks
such as Nissenbaum's theory of Contextual Integrity, 
that may offer a much needed counterpoint and additional tool of governance.

\bibliographystyle{plain}
\bibliography{ra.bib}

\end{document}